\documentclass{article}
\usepackage{amsmath, amssymb, amsfonts}
\title{ON THE HAWKING WORMHOLE HORIZON ENTROPY}
\author{Hristu Culetu, \\Ovidius University, Dept.of Physics, \\B-dul Mamaia 124, 8700 Constanta, Romania \\e-mail : hculetu@yahoo.com}

\begin{document}
\numberwithin{equation} {section}
\pagenumbering{arabic}
\maketitle

\newcommand{\fv}{\boldsymbol{f}}
\newcommand{\tv}{\boldsymbol{t}}
\newcommand{\gv}{\boldsymbol{g}}
\newcommand{\OV}{\boldsymbol{O}}
\newcommand{\wv}{\boldsymbol{w}}
\newcommand{\WV}{\boldsymbol{W}}
\newcommand{\NV}{\boldsymbol{N}}
\newcommand{\hv}{\boldsymbol{h}}
\newcommand{\yv}{\boldsymbol{y}}
\newcommand{\RE}{\textrm{Re}}
\newcommand{\IM}{\textrm{Im}}
\newcommand{\rot}{\textrm{rot}}
\newcommand{\dv}{\boldsymbol{d}}
\newcommand{\grad}{\textrm{grad}}
\newcommand{\Tr}{\textrm{Tr}}
\newcommand{\ua}{\uparrow}
\newcommand{\da}{\downarrow}
\newcommand{\ct}{\textrm{const}}
\newcommand{\xv}{\boldsymbol{x}}
\newcommand{\mv}{\boldsymbol{m}}
\newcommand{\rv}{\boldsymbol{r}}
\newcommand{\kv}{\boldsymbol{k}}
\newcommand{\VE}{\boldsymbol{V}}
\newcommand{\sv}{\boldsymbol{s}}
\newcommand{\RV}{\boldsymbol{R}}
\newcommand{\pv}{\boldsymbol{p}}
\newcommand{\PV}{\boldsymbol{P}}
\newcommand{\EV}{\boldsymbol{E}}
\newcommand{\DV}{\boldsymbol{D}}
\newcommand{\BV}{\boldsymbol{B}}
\newcommand{\HV}{\boldsymbol{H}}
\newcommand{\MV}{\boldsymbol{M}}
\newcommand{\be}{\begin{equation}}
\newcommand{\ee}{\end{equation}}
\newcommand{\ba}{\begin{eqnarray}}
\newcommand{\ea}{\end{eqnarray}}
\newcommand{\bq}{\begin{eqnarray*}}
\newcommand{\eq}{\end{eqnarray*}}
\newcommand{\pa}{\partial}
\newcommand{\f}{\frac}
\newcommand{\FV}{\boldsymbol{F}}
\newcommand{\ve}{\boldsymbol{v}}
\newcommand{\AV}{\boldsymbol{A}}
\newcommand{\jv}{\boldsymbol{j}}
\newcommand{\LV}{\boldsymbol{L}}
\newcommand{\SV}{\boldsymbol{S}}
\newcommand{\av}{\boldsymbol{a}}
\newcommand{\qv}{\boldsymbol{q}}
\newcommand{\QV}{\boldsymbol{Q}}
\newcommand{\ev}{\boldsymbol{e}}
\newcommand{\uv}{\boldsymbol{u}}
\newcommand{\KV}{\boldsymbol{K}}
\newcommand{\ro}{\boldsymbol{\rho}}
\newcommand{\si}{\boldsymbol{\sigma}}
\newcommand{\thv}{\boldsymbol{\theta}}
\newcommand{\bv}{\boldsymbol{b}}
\newcommand{\JV}{\boldsymbol{J}}
\newcommand{\nv}{\boldsymbol{n}}
\newcommand{\lv}{\boldsymbol{l}}
\newcommand{\om}{\boldsymbol{\omega}}
\newcommand{\Om}{\boldsymbol{\Omega}}
\newcommand{\Piv}{\boldsymbol{\Pi}}
\newcommand{\UV}{\boldsymbol{U}}
\newcommand{\iv}{\boldsymbol{i}}
\newcommand{\nuv}{\boldsymbol{\nu}}
\newcommand{\muv}{\boldsymbol{\mu}}
\newcommand{\lm}{\boldsymbol{\lambda}}
\newcommand{\Lm}{\boldsymbol{\Lambda}}
\newcommand{\opsi}{\overline{\psi}}
\renewcommand{\tan}{\textrm{tg}}
\renewcommand{\cot}{\textrm{ctg}}
\renewcommand{\sinh}{\textrm{sh}}
\renewcommand{\cosh}{\textrm{ch}}
\renewcommand{\tanh}{\textrm{th}}
\renewcommand{\coth}{\textrm{cth}}

\begin{abstract}

~~The entropy S of the horizon $\theta = \pi/2$ of the Hawking wormhole written in spherical Rindler coordinates is computed in this letter.~Using Padmanabhan's prescription, we found that the surface gravity of the horizon is constant and equals the proper acceleration of the Rindler observer. S is a monotonic function of the radial coordinate $\xi$ and vanishes when $\xi$ equals the Planck length. In addition, its expression is similar with the Kaul-Majumdar one for the black hole entropy, including logarithmic corrections in quantum gravity scenarios.
\end{abstract}

~Keywords : surface gravity, spherical Rindler coordinates, horizon entropy.


\section{\textbf{Introduction}}
~The connection between gravity and thermodynamics is one of the most surprising features of gravity. Once the geometrical meaning of gravity is accepted, surfaces which act as one-way membranes for information will arise, leading to some connection with entropy, interpreted as the lack of information \cite {TP1} \cite{TP2}.

As T.Padmanabhan has noticed \cite{TP3}, in any spacetime we might have a family of observers following a congruence of timelike curves which have no access to part of spacetime (a horizon is formed which blocks the informations from those observers). 
\\Keeping in mind that QFT does not recognize any nontrivial geometry of spacetime in a local inertial frame, we could use a uniformly accelerated frame (a local Rindler frame) to study the connection between one way membranes arising in a spacetime and the thermodynamical entropy.

Another principle to which the horizon entropy is strongly related is the Holographic Principle (\cite{RB} and references therein) which states that the number of degrees of freedom describing the physics inside a volume - including gravitation - is bounded by the area of the boundary which encloses the volume.\\ As entropy counts the microscopical degrees of freedom of a physical system, it can be shown that \cite{RB}\cite{GH}\cite{DM}
\begin{equation}
S \leq \frac{A}{4}
\label{1.1}
\end{equation}
where A is the boundary area. S equals $A/4$ only for a spacetime with a horizon (black hole horizon, de Sitter cosmological horizon, Rindler horizon, etc.).

\section{\textbf{The Ricci scalar splitting}}
~~The purpose of the present letter is to compute the horizon entropy for the (Lorentzian version) of the Hawking wormhole entropy \cite {HC1}, written in (static) spherical Rindler coordinates.
\begin{equation}
ds^2 = \left(1-\frac{b^{2}}{\xi^{2}}\right)^{2} \left(-g^{2} \xi^{2}cos^{2} \theta dt^{2}+d\xi^{2} + \xi^{2} d\Omega^{2}\right)
\label{2.1}
\end{equation}
\\The spacetime (2.1) may be obtained from the Hawking wormhole metric written in Cartesian coordinates \cite{HC2}
\begin{equation}
ds^{2} = \left(1-\frac{b^{2}}{x^{\alpha}x_{\alpha}}\right)^{2} \eta_{\mu \nu} dx^{\mu} dx^{\nu}
\label{2.2}
\end{equation}
by means of the coordinate transformation
\begin{equation}
\begin{split}
x^{1} = \xi ~sin \theta~ cos \phi,~~x^{2} = \xi~ sin\theta~ sin\phi,\\
x^{3} = \xi~ cos\theta~ cosh gt,~~x^{0} = \xi~ cos\theta~ sinh gt.
\end{split}
\label{2.3}
\end{equation}
$x^{\alpha}$ $(\alpha = 0, 1, 1, 3)$ are the Minkowski coordinates, $(t, \xi, \theta, \phi)$ - the spherical Rindler coordinates, b - the wormhole's throat radius (which will be taken of the order of the Planck length), $d\Omega^{2} = d\theta^{2} + sin^{2}\theta d\phi^{2}$, g - a constant with units of acceleration, $\eta_{\mu \nu} = diag~(-1, 1, 1, 1)$ and $x^{\alpha} x_{\alpha} = \xv^{2} - \left(x^{0}\right)^{2}$.
\\The spacetime (2.1) has a horizon at $\xi = b$, which is also a null geodesic (the hypersurface $\xi = b$ is in fact the Hawking wormhole which separates the two causally disconnected, asymptotically flat regions, $\xi >> b$ and $\xi << b$). The hypersurface $\xi = 0$ from the Rindler geometry ($b = 0$) is no longer a horizon here due to the conformal factor).

From now on we take into consideration only the region $\xi > b$. The units will be such that $c = G = \hbar = k_{B} = 1$.

Let us now use Padmanabhan's prescription \cite {TP2} to calculate the entropy of the horizon $\theta = \pi/2$, where the time-time component of the metric (2.1) is vanishing (the hemispheres $0\leq \theta <\pi/2$ and $\pi/2 <\theta \leq\pi$ correspond to the two Rindler observers which are causally disconnected).
\\One can show \cite {TP2} that
\begin{equation}
\begin{split}
R = ^{3}R+K_{\alpha \beta} K^{\alpha \beta}-\left(K^{\alpha}_{\alpha}\right)^{2}+2~\nabla_{\alpha}(K u^{\alpha}+a^{\alpha}) \equiv \\
L+2~\nabla_{\alpha}(K u^{\alpha}+a^{\alpha})
\end{split}
\label{2.4}
\end{equation}
where R is the 4-dimensional Ricci scalar, $^{3}R$ - the scalar curvature of a spacelike hypersurface $\Sigma$ with $u^{\alpha} (\alpha = 0, 1, 2, 3)$ as normal, $K_{\alpha \beta}$ is the extrinsic curvature of $\Sigma$ with $K = K^{\alpha}_{\alpha}$, $a^{\alpha} = u^{\beta}~\nabla_{\beta} u^{\alpha}$ - the corresponding acceleration and L - the ADM Lagrangean. We integrate eq.(2.4) over a four-volume $\Omega$, bounded by $\Sigma$ and by a timelike surface B, with normal $n^{\alpha}$ \cite{TP2}. The induced metric on $\Sigma$ is $h_{\mu \nu} = g_{\mu \nu}+u_{\mu} u_{\nu}$ and the metric on B is $\gamma_{\mu \nu} = g_{\mu \nu}-n_{\mu} n_{\nu}$. The hypersurfaces $\Sigma$ and B intersect on a two-dimensional surface $\Sigma \cap B$ on which the geometry is 
\begin{equation}
\sigma_{\alpha \beta} = g_{\alpha \beta}+u_{\alpha} u_{\beta}-n_{\alpha} n_{\beta}
\label{2.5}
\end{equation}
We observe that the metric (2.1) can be put in the form \cite {TP3}
\begin{equation}
ds^{2} = -N^{2}(\xv) dt^{2}+f_{ij}(\xv) dx^{i} dx^{j} 
\label{2.6}
\end{equation}
where $i, j = 1, 2, 3$. In our case 
\begin{equation}
N(\xv) = g \xi \left(1-\frac{b^{2}}{\xi^{2}}\right) cos~\theta ,
\label{2.7}
\end{equation}
Taking $\Sigma$ to be the surface of constant time and keeping in mind that the metric (2.1) is static, the trace K is vanishing. By integration of $R/16 \pi$ over $\Omega$, the term with L will give the ADM energy. 

\section{\textbf{The horizon entropy}}
~~Let us consider B as the surface $\theta = \pi/2$, the horizon obtained from the condition $N = 0$. The last term in the r.h.s. of (2.4) may be transformed in a surface integral over B, giving the entropy of the horizon
\begin{equation}
S = \frac{1}{8 \pi} \int_{B} a_{\alpha} n^{\alpha} N \sqrt{\sigma} d\xi~ d\phi~ dt ,
\label{3.1}
\end{equation}
where $\sigma$ is the determinant of the metric on the two-surface $\Sigma \cap B$. Since the acceleration vector $a_{\alpha}$ is spacelike, we can put $a_{0} = 0$ at a given event in a local Rindler frame \cite{TP3}. Therefore
\begin{equation}
a_{\alpha} \equiv \frac{N_{\alpha}}{N} = \left(0,~\xi^{-1}(1-\frac{b^{2}}{\xi^{2}})^{-1} (1+\frac{b^{2}}{\xi^{2}}),~-tg~\theta,~0\right)
\label{3.2}
\end{equation}
As the surface B approaches the horizon, the expression $N~a^{\mu}n_{\mu}$ tends to the surface gravity $\kappa$ of the hotizon $\theta = \pi/2$. In our static spacetime (2.1) with a horizon, the Euclidian action will be periodic in imaginary time with the period $T = 2 \pi/g$. In this case $t \in (0, T)$.

The normal vector to the hypersurfaces $\Sigma$ and B appears as 
\begin{equation}
u_{\alpha} = \left[~g~\xi~(1-\frac{b^{2}}{\xi^{2}})~cos~\theta,~0,~0,~0\right] 
\label{3.3}
\end{equation}
and, respectively
\begin{equation}
n_{\alpha} = \left[0,~0,~\xi~(1-\frac{b^{2}}{\xi^{2}}),~0)\right]
\label{3.4}
\end{equation}
The corresponding metrics become
\begin{equation}
ds^{2}|_{\Sigma} = \left(1-\frac{b^{2}}{\xi^{2}}\right)^{2}~(d~\xi^{2}+\xi^{2}~d~\Omega^{2})
\label{3.5}
\end{equation}
and
\begin{equation}
ds^{2}|_{B} = \left(1-\frac{b^{2}}{\xi^{2}}\right)^{2}~(-g^{2}~\xi^{2}~cos^{2}\theta~dt^{2}+d~\xi^{2}+\xi^{2}~sin^{2}\theta~d\phi^{2})
\label{3.6}
\end{equation}
while the geometry on the two - surface $\Sigma \cap B$ acquires the form
\begin{equation}
d\sigma^{2} = \left(1-\frac{b^{2}}{\xi^{2}}\right)^{2}~(d\xi^{2}+\xi^{2}sin^{2}\theta~d\phi^{2})
\label{3.7}
\end{equation}
The surface gravity $\kappa$ will be given by
\begin{equation}
\begin{split}
\kappa = Na_{\alpha}n^{\alpha}|_{\theta = \pi/2} = \sqrt{g^{\alpha \beta}N_{,\alpha}N_{,\beta}}~|_{\theta=\pi/2} = \\g~(1-\frac{b^{2}}{\xi^{2}})^{-1}~\sqrt{\left(1+\frac{b^{2}}{\xi^{2}}\right)^{2}~cos^{2}\theta+\left(1-\frac{b^{2}}{\xi^{2}}\right)^{2}~sin^{2}\theta}~|_{\theta = \pi/2} = g.
\end{split}
\label{3.8}
\end{equation}
In other words, the meaning of the constant g is just the surface gravity of the horizon $\theta = \pi/2$. Therefore, the expression of the entropy S will be given by
\begin{equation}
S = \frac{g}{8 \pi}~\int_{0}^{T}~dt~\int_{0}^{2 \pi}~\int_{0}^{\xi}~\sqrt{\sigma}~d\xi d\phi
\label{3.9}
\end{equation}
After an integration over the imaginary time, the entropy of the horizon $\theta = \pi/2$ is given by the well known expression in terms of the horizon area.

It would be interesting to find the function $S(\xi)$. Keeping in mind that
\begin{equation}
\sqrt{\sigma} = \xi \left(1-\frac{b^{2}}{\xi^{2}}\right)^{2}~sin^{2}\theta,
\label{3.10}
\end{equation}
we have from (3.8) (with $\theta = \pi/2$)
\begin{equation}
S(\xi) = \frac{\pi \xi^{2}}{4 b^{2}}~\left(1+\frac{4 b^{2}}{\xi^{2}}\ln \frac{b}{\xi}-\frac{b^{4}}{\xi^{4}}\right)
\label{3.11}
\end{equation}
where $l^{2}_{P} = b^{2}$ was introduced at the denominator.
 It is an easy task to show that $S(\xi)$ is a monotonic function. It vanishes at $\xi = b$ and, for $\xi >> b$, S increases to the value $\pi \xi^{2}/4 b^{2} \equiv a/4b^{2}$ ($a$ represents the area of a circle of radius $\xi$ - a part of the full horizon).

Let us compare (3.11) with the expression for a black hole entropy including logarithmic corrections in quantum gravity (QG) scenarios \cite{DF} \cite{MA} \cite{KM}
\begin{equation}
S_{QG} = a/4l_{P}^{2} + \alpha ln\frac{a}{l_{P}^{2}} + O(\frac{l_{P}^{2}}{a})
\label{3.12}
\end{equation}
Eq.(3.12) is derived by a direct microstate counting in string theory and loop quantum gravity when the coefficient $\alpha$ - which depends on the number of field species - is negative. Writing eq. (3.11) in the form
\begin{equation}
 S(\xi) = \frac{a(\xi)}{4l_{P}^{2}} - \frac{\pi}{2} ln\frac{a}{b^{2}} + \pi ln2 - \frac{\pi b^{2}}{a}
 \label{3.13}
 \end{equation}
 We observe the similarity with the Kaul - Majumdar expression (3.12), with $\alpha = -\pi/2$ in our case.
 
\section{\textbf{Conclusions}}

We applied in this paper Padmanabhan's method to compute the entropy of the horizon $\theta = \pi/2$ for the Hawking wormhole spacetime, written in (static) spherical Rindler coordinates. The surface gravity is constant (note that the lapse function N depends on two variables, $\xi$ and $\theta$). In addition, the Hawking temperature of the horizon is given by $g/2 \pi$, since $\kappa = g$. The entropy is a monotonic function ; it increases from zero at $\xi = b$ to $\pi \xi^{2}/4 b^{2}$ at $\xi >> b$ (note that $\xi^{2}$ is just the Minkowski interval).

A comparison between our result (3.11) and the corrected Kaul - Majumdar form of the Bekenstein - Hawking black hole entropy leads to the conclusion that they have similar structures, only the coefficients being different.

\textbf{Acknowledgments}

This paper was presented at the 2-nd Vienna Central European Seminar "`Frontiers in Astroparticle Physics"', Vienna, November 25 - 27, 2005.

\end{document}